\documentclass{article}
\usepackage[utf8]{inputenc}
\usepackage[english]{babel}
\usepackage[affil-it]{authblk}
\usepackage[margin=0.8in]{geometry}
\usepackage{graphicx}
\usepackage[noadjust]{cite}
\usepackage{amsmath}
\usepackage{amssymb}

\begin{document}

\title{Testing the Equivalence Principle with Unstable Particles}

\author[1,2]{Y. Bonder\footnote{email: ybonder@indiana.edu}}
\author[3]{E. Fischbach}
\author[1,4]{H. Hernandez-Coronado}
\author[3,5]{D.E. Krause}
\author[5]{Z. Rohrbach}
\author[1]{D. Sudarsky}

\affil[1]{Instituto de Ciencias Nucleares, Universidad Nacional Aut\'onoma de M\'exico, Apartado Postal 70-543, M\'exico D.F., 04510, M\'exico.}
\affil[2]{Physics Department, Indiana University, Bloomington, Indiana, 47405, USA.}
\affil[3]{Department of Physics, Purdue University, West Lafayette, Indiana, 47907, USA.}
\affil[4]{Instituto Mexicano del Petr\'oleo, Eje Central L\'azaro C\'ardenas 152, M\'exico D.F., 07730, M\'exico.}
\affil[5]{Physics Department, Wabash College, Crawfordsville, Indiana, 47933, USA.}

\date{(Dated: IUHET 576, May 2013)}

\maketitle
\begin{abstract}
We develop a framework to test the Equivalence Principle (EP) under conditions where the quantum aspects of nature cannot be neglected, specifically in the context of interference phenomena with unstable particles. We derive the nonrelativistic quantum equation that describes the evolution of the wavefunction of unstable particles under the assumption of the validity of the EP and when small deviations are assumed to occur. As an example, we study the propagation of unstable particles in a COW experiment, and we briefly discuss the experimental implications of our formalism.
\end{abstract}

\section{Motivation}

The Equivalence Principle (EP) plays a central role in our understanding of nature. It lies at the basis not only of the general theory of relativity, but of the notion of inertia itself, since the practical realization of an inertial frame is only possible by assuming the EP. It is thus not surprising that testing such a principle should have a tradition that is almost as old as modern physics. Experimental programs have sought to explore its validity for the most diverse kinds of physical systems and conditions \cite{EP-tests1,EP-tests2,EP-tests3,EP-tests4,EP-tests5,EP-tests6,EP-tests7,EP-tests8} ranging from space-bound experiments to tests using antimatter \cite{EP-antimatter1,EP-antimatter2,EP-antimatter3,EP-antimatter4}. One of the aspects where the principle has been least explored, perhaps due to technical difficulties, is the quantum realm, that is, those situations where the quantum/gravity interface becomes an essential aspect of the experiment. Among the most conspicuously quantum phenomena are 
those associated with unstable particles, and hence the focus of this paper will be tests of the EP with unstable quantum systems.

Among the limited instances where the quantum/gravity interface has been explored are the well-known COW experiments \cite{COW} and the recent cold neutron experiments \cite{Cold-Neutron}. Their results, which are in accordance with our theoretical expectations, have led to a degree of confusion resulting from an imprecise statement of the principle one wishes to test \cite{Dharam, Chryss, Knox}. In the classical context one of the simplest versions of the EP is the universality of free fall. This is often stated as the independence of the acceleration of test objects on their initial velocity, mass or composition \cite{UFF1,UFF2,UFF3,UFF4,UFF5,UFF6,UFF7,UFF8,UFF9,UFF10,UFF11}, \textit{i.e.}, the object's acceleration should depend only on its location. In the quantum context it is clear that the terms must be modified, not only because the objects cannot have precise position and initial velocity, but because, even in the absence of gravity, the mass of a particle also determines how the wavepacket spreads 
(see also Refs. \cite{MassQM1,MassQM2}).

A recent discussion of such issues can be found in Ref. \cite{Okon} where four broad categories of the EP are considered, two of which are suitable for applicability to the quantum context. The difference between the two is that one refers to a situation where the gravitational field can be considered homogeneous (isotropic and time independent), and the other involves deviations from that. The version we use is the first which can be stated as follows: \textit{Given a physical system contained in a spacetime region with a constant gravitational field, the description of its behavior is equivalent to that of the same system placed in a spacetime region where gravity is absent but subjected to a constant acceleration of corresponding magnitude.}

The exact nature of this equivalence is not always completely trivial. For instance when the physical system in question involves interaction with an external nongravitational field, as in the case of an accelerated charged particle. In those cases, one has to realize that such field cannot be considered as contained in the specified region and the appropriate quantum state of the field must be carefully identified \cite{Bremstrahlung1,Bremstrahlung2,Bremstrahlung3,Bremstrahlung4,Bremstrahlung5,Bremstrahlung6}.

Returning to unstable systems, we note that one central characteristic of such systems is that they cannot be eigenstates of the Hamiltonian and, as such, they have an intrinsic energy uncertainty inversely proportional to the mean lifetime of the system. This is reflected in the fact that unstable particles are characterized by a complex mass $\tilde{m}=m-i\Gamma/2$, where the imaginary part is a measure of the particle's inverse lifetime. Thus, testing the EP with unstable objects amounts to exploring how gravity affects the intrinsic energy indefiniteness of such systems. Note that even perfectly stable systems might be in states that are not energy eigenstates (\textit{e.g.}, a free particle in a highly localized state), and we expect gravity to affect all types of energy uncertainties in equal fashion. However, this expectation relies on the very principle we want to test.

The issue can be characterized at the phenomenological level as to whether the real and imaginary parts of the mass couple to gravity in accordance with the EP. (Note that we are not assuming that they should couple in the same form.) As we shall see, the question of what the precise relationship should be between the couplings of gravity to the real and imaginary parts of the mass, assuming the EP, is a rather nontrivial one. This is particularly relevant here because we are interested in the nonrelativistic equation describing unstable particles in the presence of gravity where notions such as potential energy are well defined. We note that even in the absence of gravity it is not completely clear how to obtain the nonrelativistic equation describing unstable particles. In fact, straightforward approaches to this question such as using the Schr\"odinger equation that results from replacing all masses by the complex mass are problematic. One issue is that, although the resulting equation seems Galilean 
invariant, the probability density is in conflict with this symmetry. This problem is closely related to Bargmann's superselection rule \cite{Hector}. These arguments indicate that it is not possible to simply guess how the complex mass should enter into the gravitational potential term of the evolution equation. The only way to determine the exact form of such a term is to start from a trusted basic framework which is fully compatible with the fundamental principles we wish to explore. Thus we assume as our stating point quantum field theory\footnote{Unstable particles can only be formally treated in the framework of quantum field theory where the field's self-energy, which accounts for the dressing of the particle by the interactions, modifies the simple poles present in the free theory. That, in turn, affects the wavepacket propagation in two ways: The appearance of branch cuts in the propagator, which leads to a well known power law decay of the wavepackets \cite{Nonexponetial-decay1,Nonexponetial-decay2}, and the displacement of the simple poles into the complex plane. In this work we restrict our attention to situations where the nonexponential decays associated with the branch cuts can be neglected and we focus on the poles shift which can be characterized by letting the mass become complex.}, and then carefully take the nonrelativistic limit. In the process of obtaining the desired equation, we will deal with some subtleties that could be relevant in other contexts.

\section{Formalism}

\subsection{Starting point}

The appropriate fundamental framework for the problem involving unstable particles is quantum field theory. In fact, strictly speaking, unstable particles should only be considered as virtual particles characterized by a their dressed propagator \cite{Sachs}. The experimental question has to be posed in terms of the probabilities of detecting certain out-states with corresponding localized wavepackets, having prepared the in-states as suitable localized wavepackets. The issue of spacetime evolution is then transformed into a question regarding the dependence of the probability amplitudes on the parameters describing spacetime localization of those wavepackets. We do not reproduce all those arguments here and refer the reader to Ref. \cite{Sachs} and subsequent modifications \cite{Thesis}. Instead, we start our analysis with the practical conclusion of those works which show that, to a very high degree of precision, the situations of interest can be described by the usual relativistic wave equation for 
the unstable particles with a replacement of the mass by a complex mass $\tilde{m}$. For simplicity, we work with scalar particles $\phi$ and we use units where $c=\hbar=1$, in which case the wave equation is
\begin{equation}
0=g^{\mu\nu}\nabla_\mu \nabla_\nu \phi+\tilde{m}^2\phi,
\end{equation}
where $g^{\mu\nu}$ is the (inverse of the) spacetime metric and $\nabla_\mu$ is its associated covariant derivative. This equation can be written in terms of ordinary derivatives as
\begin{equation}\label{unstable KG grav}
0=g^{\mu\nu}\partial_\mu \partial_\nu \phi-g^{\mu\nu}\Gamma^\rho_{\mu\nu}\partial_\rho \phi+\tilde{m}^2\phi,
\end{equation}
where $\Gamma^\rho_{\mu\nu}$ are the Christoffel symbols. We note that at this level there is no ambiguity of which mass should be complex because there is only one mass in Eq.~(\ref{unstable KG grav}). In order to take the nonrelativistic limit of this equation, we first need to restrict Eq.~(\ref{unstable KG grav}) to the framework of single particle relativistic quantum mechanics where $\phi$ is taken to be a wavefunction and not a quantum field; the context where the probabilistic interpretation of $\phi$ is justified is discussed below.

Assuming that the EP holds, we can introduce a uniform gravitational field by using the metric of the flat spacetime associated with a frame which has constant acceleration $\vec{a}$. Following Ref. \cite{Hehl}, this metric can be written as
\begin{equation}
ds^2=(1+\vec{a}\cdot \vec{x})^2dt^2-d\vec{x}^2,
\end{equation}
where the standard notation for $3$-dimensional vectors and Euclidean scalar product are used. As the EP is assumed to hold at this stage, we can replace $\vec{a}\cdot \vec{x}$ by a uniform Newtonian gravitational potential which we denote by $U(\vec{x})$. Note that $U$ is dimensionless and $U\geq 0$. With this replacement Eq.~(\ref{unstable KG grav}) can be written as
\begin{equation}\label{unstable KG grav 2}
0=\frac{\ddot{\phi}}{(1+U)^2}-\mathbf{\nabla}^2\phi-\frac{\mathbf{\nabla} U\cdot \mathbf{\nabla}\phi}{1+U}+\tilde{m}^2 \phi,
\end{equation}
where the overdots on $\phi$ denote time derivatives, and we again use the Euclidean notation for the $3$-dimensional gradient and Laplacian. We turn now to obtain the nonrelativistic limit of this equation.

\subsection{Nonrelativistic limit}

When the particle is free and the mass is real, the nonrelativistic limit of the Klein-Gordon equation can be studied in terms of Fourier transformations of $\phi$ in the space and time variables\footnote{The nonrelativistic limit of the Klein-Gordon equation with real mass in noninertial frames presents some additional subtleties which are discussed in Ref.~\cite{Padmanabhan}.}. The Fourier variables associated with $t$ and $\vec{x}$ can be interpreted as those obtained by performing the corresponding inverse Fourier transformations. In the present case, where $U\neq 0$ and where the mass is complex, the situation is much more delicate. To begin with, the gravitational potential depends on $\vec{x}$ so a Fourier transform would depend on momentum derivatives. This issue requires some care but, in principle, could be overcome. A more difficult problem is that the Fourier variables are real by definition. However, when dealing with a complex mass, $E$ and/or $\vec{p}$ must acquire an imaginary part in 
order to solve the wave equation, thus forcing us to leave the domain of standard Fourier transforms. One possibility is to Fourier transform our solution only with respect $t$ or $\vec{x}$, but not both. Even this approach raises further questions, for instance: When should we Fourier transform with respect to $t$ and when with respect to $\vec{x}$? How can we take the nonrelativistic limit without a standard (\textit{i.e.}, real) dispersion relation? And, what is the physical meaning of the solutions to the nonrelativistic equation?

The answer to the first question is connected with the specific experiment one is describing. It turns out that if the experimental situation is such that the wavefunction is known at a given point $\vec{x}$ for all $t$, namely, if it is described in terms of \emph{boundary} conditions, then it is convenient to Fourier transform with respect to time, namely,
\begin{equation}\label{Fourier phi}
\phi(t,\vec{x})=\frac{1}{\sqrt{2\pi}}\int dE e^{-i E t}\varphi_E(\vec{x}),
\end{equation}
where, $E$ is real and is integrated from $-\infty$ to $\infty$. The important point is that, being real, $E$ can be interpreted as the energy. On the other hand, when \emph{initial} conditions are given, \textit{i.e.}, when the wavefunction is known everywhere in space at a particular time $t$, it is convenient to use a Fourier transform with respect to $\vec{x}$, and then $\vec{p}$, which is real, can be thought as the momentum. Note that situations that are not associated with either of these experimental conditions are probably best treated with a different method. In what follows we consider the case where the wavefunction information is presented in the form of boundary conditions. If we were instead interested in experiments characterized in terms of initial conditions the corresponding nonrelativistic equation can be obtained \textit{mutatis mutandis}. We should emphasize that there is no reason to expect that the nonrelativistic equations obtained in both cases coincide, as we now discuss.

In order to obtain the nonrelativistic limit of Eq.~(\ref{unstable KG grav 2}) without an explicit dispersion relation, we must first understand what is the meaning of a limit of a differential equation. We address this question by identifying a differential equation with the space of all of its solutions $S$. Next, we note that a limiting procedure, such as the one we must confront, only makes sense in connection with a selection of a particular subset $A\subset S$ which characterizes the regime we are interested in. Assuming one has made an adequate selection of $A$, we refer to the limiting equation as the equation whose space of solutions coincides with $A$ at a certain predetermined degree of accuracy. Clearly, a different selection of $A$ would generically lead to a different limiting equation. Moreover, due to the fact that a finite degree of accuracy is involved, there would be, in principle, various equations that satisfy the above requirement. Thus, in general there would be no equation that could 
be considered as \emph{the} unique limit of the original equation.

Having clarified our approach to this issue, we now proceed to characterize the subset of nonrelativistic solutions corresponding to an experiment with unstable particles, in terms of the appropriate boundary conditions. In order to do so, we first need to find the \emph{full} set of solutions. This is done by requiring $\phi$, as given in Eq.~(\ref{Fourier phi}), to be a solution of Eq.~(\ref{unstable KG grav 2}), which implies
\begin{eqnarray}
0&=&\frac{1}{\sqrt{2\pi}}\int dE e^{-i E t}\left(\frac{-\tilde{k}_E^2-\tilde{m}^2}{(1+U)^2}\varphi_E-\mathbf{\nabla}^2\varphi_E-\frac{\mathbf{\nabla} U\cdot \mathbf{\nabla}\varphi_E}{1+U}+\tilde{m}^2 \varphi_E\right),\label{KG fourier}
\end{eqnarray}
where, we have defined
\begin{equation}\label{def ktilde}
\tilde{k}_E^2 \equiv E^2-\tilde{m}^2.
\end{equation}
It follows that the Eq.~(\ref{KG fourier}) is satisfied, if and only if
\begin{eqnarray}
\tilde{k}_E^2\varphi_E&=& -(1+U)^2\mathbf{\nabla}^2\varphi_E-(1+U)\mathbf{\nabla} U\cdot \mathbf{\nabla}\varphi_E +\tilde{m}^2\left(2U+U^2\right)\varphi_E.\label{eq varphi}
\end{eqnarray}
Thus, a generic solution of Eq.~(\ref{unstable KG grav 2}) has the form of Eq.~(\ref{Fourier phi}) with $\varphi_E$ satisfying Eq.~(\ref{eq varphi}).

We define the nonrelativistic solutions as the subset characterized by the fact that $\varphi_E$ has support when $E$ is in the interval $[m,m+K]$, where $K\ll m$. Again, the nonrelativistic differential equation we are looking for is defined as an equation whose solution space coincides with the space of these nonrelativistic solutions; we now turn to find this equation. Let ${k}_E$ be the real part of $\tilde{k}_E$. The condition that $E$ is real translates to the fact that $E^2=\tilde{m}^2+\tilde{k}_E^2 $ is real, which in turn implies that the imaginary part of $\tilde{k}_E$ must be $m \Gamma/2 k_E$. Taking this into account, we can write
\begin{equation}\label{exact disp rel}
E=\sqrt{\tilde{m}^2+\tilde{k}_E^2}=m\sqrt{1+\frac{k_E^2}{m^2}}\sqrt{1-\frac{ \Gamma^2}{4 k_E^2}}.
\end{equation}
Writing $E$ in this form allows us to show that the assumption that it is contained in the interval defined by the nonrelativistic limit is equivalent to the conditions
\begin{equation}\label{Gamma bound}
1\gg \frac{k_E^2}{m^2}\geq \frac{\Gamma^2}{4 k_E^2}.
\end{equation}
Note that in this step we separate its real part, $m$, from $\tilde{m}$. As we shall see, this separation is what lies behind the asymmetry between the real and imaginary parts of $\tilde{m}$ in the nonrelativistic equation, an asymmetry which is not present in Eq.~(\ref{unstable KG grav 2}). Also observe that $k_E^2/m^2$ and
$\Gamma^2/k_E^2$ are real and dimensionless. With this in mind, it is
justified to expand Eq.~(\ref{exact disp rel}), yielding
\begin{equation}\label{1st order E}
\frac{E}{m}\approx 1+\frac{k_E^2}{2m^2}-\frac{ \Gamma^2}{8 k_E^2}=\frac{\tilde{m}}{m}+\frac{\tilde{k}_E^2}{2m^2},
\end{equation}
where in the first step we neglect terms of order $k_E^4/m^4$, $\Gamma^4/k_E^4$ and $(k_E^2/m^2)(\Gamma^2/k_E^2)= \Gamma^2/m^2$, and in the second step we reinsert $\tilde{m}$ and $\tilde{k}_E$.

In order to find the equation that generates the nonrelativistic solutions we calculate
\begin{eqnarray}
i\partial_t \phi&=&\frac{1}{\sqrt{2\pi}}\int dE E e^{-i E t}\varphi_E\nonumber\\
&=& \frac{1}{\sqrt{2\pi}}\int dE \left(\tilde{m}+\frac{\tilde{k}_E^2}{2m}\right) e^{-i E t}\varphi_E,\label{nonrel eq}
\end{eqnarray}
where in the last step we use the fact that in the nonrelativistic regime $\varphi_E$ has support in the region where we can use Eq.~(\ref{1st order E}). Using Eq.~(\ref{eq varphi}) we can replace $\tilde{k}_E^2\varphi_E$ by an $E$-independent operator acting on $\varphi_E$, obtaining
\begin{eqnarray}
i\partial_t \phi
&=&\frac{1}{\sqrt{2\pi}}\int dE e^{-i E t}\left\{\tilde{m}-\frac{1}{2m} \left[(1+U)^2\mathbf{\nabla}^2-(1+U)\mathbf{\nabla} U\cdot \mathbf{\nabla}\right]+\frac{\tilde{m}^2}{m} \left(U+\frac{U^2}{2}\right)\right\}\varphi_E\nonumber\\
&=&\left\{\tilde{m}-\frac{1}{2m} \left[(1+U)^2\mathbf{\nabla}^2-(1+U)\mathbf{\nabla} U\cdot \mathbf{\nabla}\right]+\frac{\tilde{m}^2}{m} \left(U+\frac{U^2}{2}\right)\right\}\phi,
\end{eqnarray}
where in the last equality we use the fact that the operator inside the braces can be removed from the integral. The important point is that we now have a Schr\"odinger equation where the coefficients appearing in the kinetic and potential terms are determined. The coefficient in the terms involving spatial derivatives of the wavefunction is simply the standard $1/2m$, with the real part of the mass, and that appearing in the purely potential terms turns out to be $\tilde{m}^2/m$ which is \emph{not} simply proportional to either $m$ or $\tilde{m}$. Observe that, at the level of approximation we are working, $\tilde{m}^2/m \approx m-i\Gamma$, which, after some rearrangements leads to 
\begin{eqnarray}\label{nonrel eq 3}
i\partial_t \phi&=&\left[m+ m U- \frac{1}{2m}(1+U)^2\mathbf{\nabla}^2 - \frac{i \Gamma }{2}-i\Gamma U+\frac{1}{2m} (1+U)\mathbf{\nabla} U\cdot \mathbf{\nabla} +(m-i\Gamma)\frac{U^2}{2} \right]\phi. 
\end{eqnarray}
The first three terms in the right-hand side of this equation can be identified as the rest energy, purely potential (gravitational) energy, and (gravitationally adjusted) kinetic energy terms. The fourth term is purely imaginary ($\Gamma$ is real and positive) and accounts for the particle's exponential decay. The fifth term involves the decay constant and the Newtonian gravitational potential and it describes the gravitational time-dilation of the decay rate, and the last terms are higher order corrections.

\subsection{Equivalence Principle violation}

The time evolution of an unstable particle in the presence of a uniform gravitational field under the assumption that the EP holds is given by Eq.~(\ref{nonrel eq 3}). Note that at this level $\Gamma$ appears in only two terms: as the standard imaginary part of the mass of a free particle, and in connection with the gravitational potential. With this equation in hand it is now straightforward to postulate an equation that could be used to parametrize violations of the EP by unstable particles.

In order to phenomenologically characterize the violations of the EP by unstable particles we add to Eq.~(\ref{nonrel eq 3}) a general complex function of $\Gamma$ and $U$ that vanishes when $\Gamma=0$ or $U=0$. To first order in $\Gamma$ and $U$, including this function is equivalent to adding a term $-i\xi \Gamma U$ where $\xi$ is a complex constant that parametrizes violations of the EP by unstable particles (the $-i$ factor is conventional). In other words, the appropriate phenomenological equation for studying possible violations of the EP by unstable particles is
\begin{eqnarray}
i\partial_t \phi&=&\left[m+ m U- \frac{1}{2m}(1+U)^2\mathbf{\nabla}^2 - \frac{i \Gamma }{2}-i(1+\xi)\Gamma U +\frac{1}{2m} (1+U)\mathbf{\nabla} U\cdot \mathbf{\nabla} +(m-i\Gamma)\frac{U^2}{2} \right]\phi. \label{nonrel eq 4}
\end{eqnarray}
where $\xi= 0$ would mean that the EP holds for unstable particles.

We emphasize that the parameter $\xi$ is purely phenomenological. As such, $\xi$ could be determined, in a more fundamental theory, by the details of the unstable system (\textit{i.e.}, its internal degrees of freedom) and the parameters of the underlying theory. For instance, when the unstable particle is an atom in an excited state, the energy levels between excited and unexcited states and the strength interaction responsible for the decay should lead to a particular value for $\xi$. However, these same details should allow us to calculate $\Gamma$. The important point is that the phenomenological term in Eq.~(\ref{nonrel eq 4}), $\xi \Gamma U$, is assumed to incorporate all these aspects and we only need to keep in mind that $\xi$ varies from one system to another.

We note that the EP-violating phenomenological term, $\xi \Gamma U$, induces violations of several related symmetries. Chief among them is Lorentz invariance, present because the phenomenological term is introduced in a particular frame. In this context we stress that our analysis has been done in the frame where the gravitational effects are simply represented by a Newtonian potential (and where, in particular, there are no gravitomagnetic effects). However, it is possible to look for alternative formulations where nontrivial aspects of the gravitational environment are considered (for an example see Refs.~\cite{QGP1,QGP2,QGP3,QGP4}). As has been shown in Ref.~\cite{FischbachA1}, a violation of Lorentz invariance can lead to a EP violation. It is thus possible that the reverse could also be true in which case unstable systems may be good candidates for experiments searching for violations of Lorentz invariance. Such violations could arise through a variety of mechanisms \cite{ReviewLVmech,FischbachA3}.

Regarding the usual conservation laws, we remark that our starting point is already an effective framework as the degrees of freedom of the decayed products are neglected. (This approximation is responsible for the lack of unitarity in the evolution represented by the standard imaginary part in the mass.) Thus, our analysis allows for a small apparent violation of energy conservation associated with the ignored degrees of freedom. On the other hand, if one applies the corresponding transformations to both the unstable particle wavefunction and the Newtonian potential in Eq.~(\ref{nonrel eq 4}), it is possible to check that our formulation is invariant under spatial translations and rotations. However, any spatial dependence of the external potential would induce apparent deviations from the momentum and angular momentum conservation of the unstable system. Moreover, as our formalism does not affect the diffeomorphism invariance in constant-time hypersurfaces, our results are covariant under spatial 
coordinate transformations in the same sense as in standard quantum mechanics.

\subsection{Physical interpretation}

Turning now to the question of the physical meaning of the nonrelativistic solutions, we note that the probability density associated with a relativistic wavefunction is given by
\begin{eqnarray}
\rho(t,\vec{x})&=&\frac{i}{2m}(\phi^* \dot{\phi}-\dot{\phi}^* \phi)\nonumber\\
&=&\frac{1}{2\pi}\int dE dE' \frac{E+ E'}{2m} \varphi_E \varphi^*_{E'} e^{-i (E-E') t},\label{rel prob}
\end{eqnarray}
where in the last step we focus on solutions that can be written in the form of Eq.~(\ref{Fourier phi}). The issue of when this probability density coincides with the one obtained from a solution to Eq.~(\ref{nonrel eq 3}) \textit{\`a la} Schr\"odinger is again related to the particular experimental conditions that are being studied. When the experiment consists of a detector placed at a fixed spatial point $\vec{x}_0$ that collects data during a long time interval (which we approximate as infinite), the measurable quantity is
\begin{eqnarray}
\int_{-\infty}^\infty dt \rho(t,\vec{x}_0)&=&\frac{1}{2\pi}\int dE dE' \frac{E+ E'}{2m} \varphi_E(\vec{x}_0) \varphi^*_{E'}(\vec{x}_0)\int_{-\infty}^\infty dt e^{-i (E-E') t} \nonumber\\
&=&\frac{1}{2\pi} \int dE dE' \frac{E+ E'}{2m} \varphi_E(\vec{x}_0) \varphi^*_{E'}(\vec{x}_0) 2\pi \delta(E-E')
\nonumber\\
&=& \int dE \frac{E}{m} |\varphi_E(\vec{x}_0)|^2.\label{IntegratedProb}
\end{eqnarray}
In the nonrelativistic limit $|\varphi_E(\vec{x}_0)|^2$ decreases faster than $\varphi_E(\vec{x}_0)$ as $E$ approaches $K$, therefore, we can take $E$ to its lowest order in the nonrelativistic expansion, namely, $E= m$. In this case,
\begin{equation}
\int_{-\infty}^\infty dt \rho(t,\vec{x}_0)= \int dE |\varphi_E(\vec{x}_0)|^2 =\int_{-\infty}^\infty dt |\phi(t,\vec{x}_0)|^2,
\end{equation}
where the last step follows from inserting $\phi$ as given by Eq.~(\ref{Fourier phi}). This fact is what ultimately allows us to interpret the solutions of Eq.~(\ref{nonrel eq 3}) as probability amplitudes in the context of nonrelativistic quantum mechanics. We note that since $\Gamma$ plays no role in this part of the analysis, one must also deal with this issue when working with stable particles. The point is that there could be situations where the relationship between the solutions of the relativistic equations and those of the corresponding limiting equation might not be as direct as the one we have found here and ensuring that this relationship is justified is paramount in preventing the introduction of spurious terms. In order to have a wavefunction interpretation, the derivation we present is not only associated with the boundary conditions described above, but also with a particular kind of measurement. (See Ref. \cite{Baute2001} for a discussion on the physical interpretation of these 
conditions.)

\section{Experimental outlook}

In this section we show how the nonrelativistic Eq.~(\ref{nonrel eq 4}) can be used to make physical predictions. To that end we calculate the propagation of a wavepacket according to this equation in the simplest interferometric experiment sensitive to gravity, namely, a COW setting \cite{COW}. Typically, to find the interference pattern one would evaluate the difference of the interaction Hamiltonian integrated over time along the two paths of the interferometer. However, in the situation at hand concepts such as proper time along a path are not well defined. This is not only due to the fact that we are dealing with quantum systems, but because the relevant states correspond to complex superpositions of components with different energies. Moreover, the effects of gravity on the evolution of each component, and the combined consequences on the full wavefunction, cannot be reliably analyzed with arguments based on our classical intuition. Thus, to be on the safe side, we make a longer calculation, presented 
in Appendix~\ref{appendix COW}, where we study wavepacket propagation in the COW setting\footnote{In Appendix~\ref{limitcases} we consider two limit cases: plane waves and extremely localized wavepackets. In the first case we recover the usual plane wave COW interference multiplied by an exponential decay. For the second case the interference term contribution to the probability vanishes as the wavepacket localization grows. This effect is caused by the relative delay of the wavepackets traveling in the two interferometer's paths. Incidentally, this result also suggests that the loss of contrast as the tilt angle of the COW interferometer increases reported in Ref. \cite{COWsystematic} could be, at least in part, due to difference in arrival times of the finite-size wavepackets.} by imposing flux conservation in each beam splitter and mirror.

\begin{figure}
\centering
\includegraphics[width=0.45\textwidth]{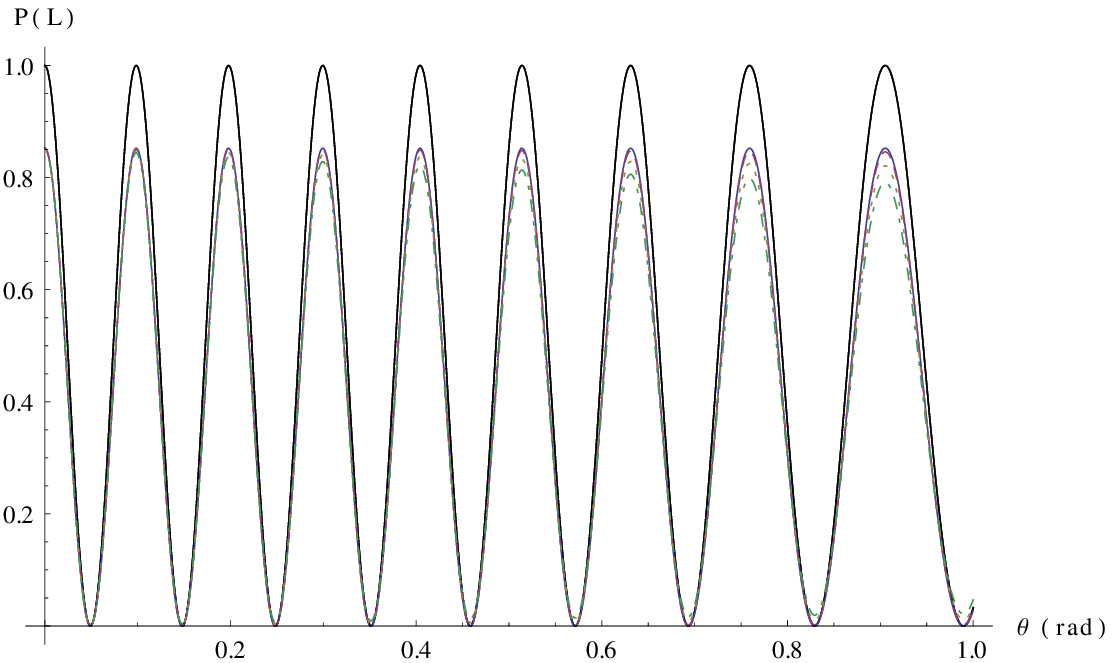}
\includegraphics[width=0.45\textwidth]{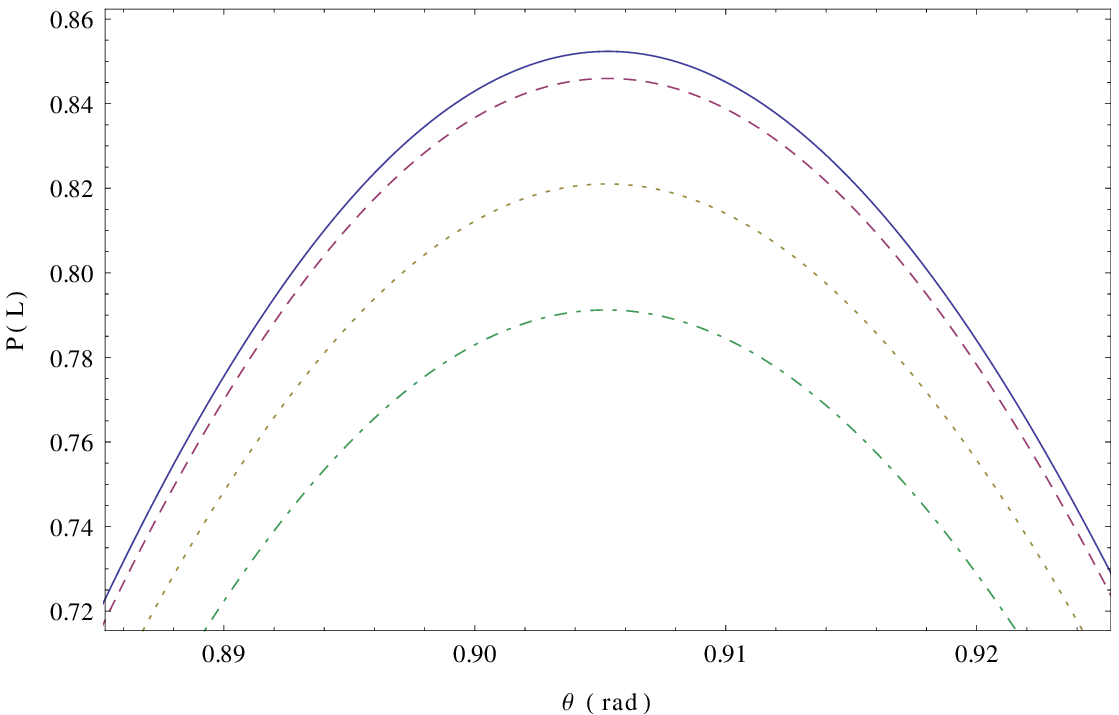}
\includegraphics[width=0.45\textwidth]{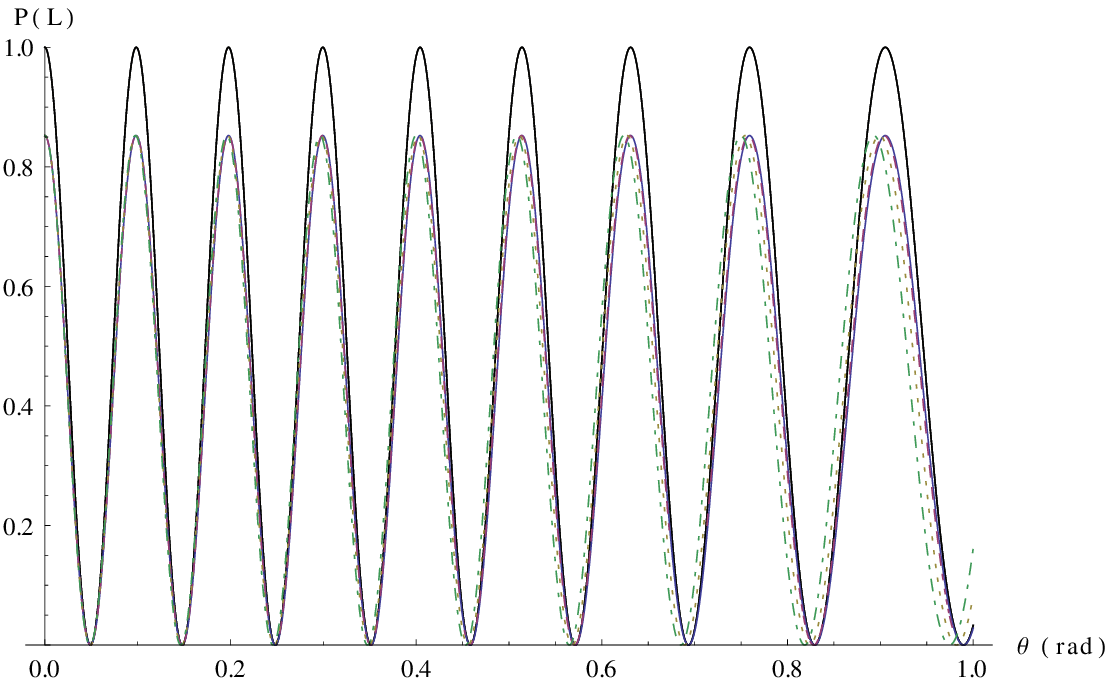}
\includegraphics[width=0.45\textwidth]{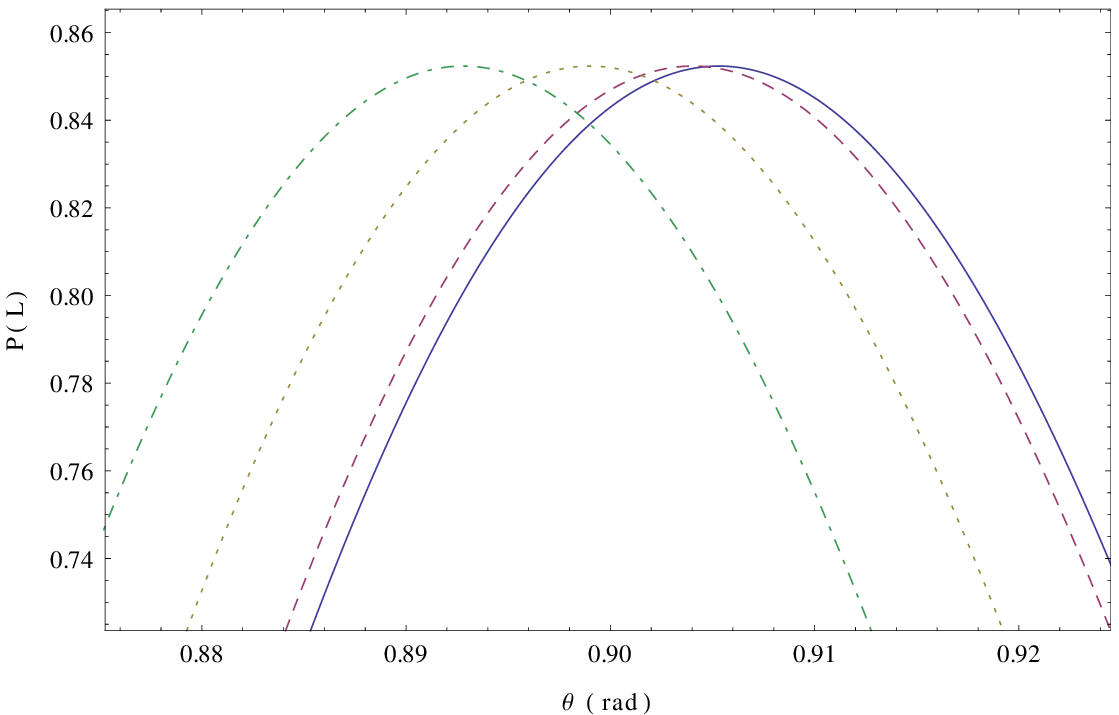}
\caption{\label{fig:fPi}The probability $P(L)$ as a function of $\theta$ for particles with initial velocity $v_0=300~\rm{m}/\rm{s}$, mass $m=939.6~\rm{MeV}$ and decay constant $\Gamma=1.5\times 10^{4}~\rm{s}^{-1}$ in a COW configuration with length and height $L=h=3.5\times 10^{-2} \rm{m}$. In all the plots, the solid (black) line corresponds to the case where $\Gamma=0$. In the top plots $\xi^I=0$ and the (blue) continuous, (magenta) dashed, (yellow) dotted and (green) dot-dashed lines correspond to $\xi^R=0,10^{16},5\times 10^{16}$ and $10^{17}$, respectively. In the bottom plots the phenomenological parameters we use are $\xi^R=0$ and $\xi^I=0,10^{17},5\times10^{17}$ and $10^{18}$ which are respectively used for the (blue) continuous, (magenta) dashed, (yellow) dotted and (green) dot-dashed lines. The plots on the right show, in greater detail, a portion of the corresponding plots on the left.}
\end{figure}

In Fig.~\ref{fig:fPi} we present the probability $P(L)$ of detecting a particle in a COW experiment, over a long time interval, as a function of the tilt angle $\theta$ between the COW apparatus and the local gravitational acceleration. We consider Gaussian energy wavepackets centered around $E_0$ whose width, $\Delta E$, is such that $\Delta E\ll E_0$. Several values of the real and imaginary parts of $\xi$ are used, which are respectively denoted by $\xi^R$ and $\xi^I$, and we take the initial velocity and the size of the experimental setting as in the original COW experiment \cite{COW}. We also assume that the particle's mass coincides with the mass of the neutrons but, for illustrative purposes, we take $\Gamma$ to be much greater than the corresponding value for neutrons.

As shown in Appendix~\ref{appendix COW}, $P(L)$ can be written as
\begin{equation}
P(L)=\int e^{-\frac{\epsilon^2}{2}}\left(e^{-\beta_u(\epsilon)}+e^{-\beta_l(\epsilon)}+2 e^{-\chi(\epsilon)}\cos{\varphi(\epsilon)}\right)d\epsilon.
\end{equation}
In the particular experimental setup we study, the terms containing $\xi^R$ in $\beta_{u,l}$ and $\chi$ are suppressed by $v_0^2 U_{u,l}$ which is of order of $10^{-28}$, while the imaginary part of $\xi$, appears suppressed by a factor $\Gamma/m$ that is of order of $10^{-20}$. Therefore, $\xi$ has to be large in order to be noticeable in the probability density plots. Moreover, the real part of $\xi$ enters only in $\beta_{u,l}$ and $\chi$, affecting only the exponential decaying factors. This is reflected in top plots in Fig.~\ref{fig:fPi} as the dependence of the peak heights in $\xi^R$. On the other hand, the imaginary part of $\xi$ only shows up in the phase $\varphi$ allowing a shift in the probability peaks position, as can also be observed in Fig.~\ref{fig:fPi}. 

From this analysis it seems that, even for extremely large values of $\xi$, it would be difficult to detect its presence, particularly since COW-like interferometers have an accuracy of the order of $1 \%$ \cite{modernCOW}. Nevertheless, the nontrivial dependence of $P(L)$ with respect to $\theta$ suggests that one could devise experiments that are more sensitive to the particular effects related with $\xi$. One interesting idea is to use the techniques developed in the atomic interference experiments of Refs.~\cite{Peters,Muller} where a remarkable accuracy of $3$ parts in $10^9$ has been achieved. 

An other possibility is to adapt experiments of the sort envisioned in Ref.~\cite{FischbachB1} for our purposes. Consider a transition from an excited state to a ground state separated by an energy $\Delta E$. Under appropriate conditions the transition matrix element could acquire a phase $~ \Gamma/\Delta E$, where $\Gamma$ is the width of the transition. For $^{181{\rm m}}_{74}\rm W$ ($T_{1/2}=14\ {\rm \mu s}$; $\Delta E = 0.37\ \rm keV$) we find $\Gamma/\Delta E\approx 10^{-13}$ which is comparable to the sensitivity of experiments considered in Ref.~\cite{FischbachB1}. Of course, new technologies would be required to deal with states having such short lifetimes. However, given the rapid advances that have been achieved in recent years, such experiments may be feasible in the foreseeable future.

We want to remark that in our derivation there is no restriction on the type of unstable particles that could be used. This fact allows us to consider any unstable particle ranging from elementary particles and hadrons to atoms in excited states and including radioactive nuclei. Still, the experimental setup could restrict the type of particles. For instance, if the time it takes the particle to traverse the interferometer is much larger than the particles lifetime, most particles will not survive the journey and will not contribute to the interference pattern, while, on the other hand, if the lifetime is much larger than the traveling time, the particles will behave mostly like stable particles.

\section{Conclusions}

In summary, we have obtained in Eq.~(\ref{nonrel eq 3}) the Schr\"odinger equation describing unstable particles in the presence of gravity by assuming that the EP is valid. In Eq.~(\ref{nonrel eq 4}) we have proposed a natural extension of Eq.~(\ref{nonrel eq 3}) which suggests how to parametrize a class of possible violations of the EP by unstable particles. In doing so, it was imperative to prevent the introduction, through the formalism itself, of aspects that might be mistakenly identified as violations of the principle that we want to test. This task was achieved by carefully taking the nonrelativistic limit of the Klein-Gordon equation for particles with complex mass as seen from a uniformly accelerated frame. As it can be seen in Eq.~(\ref{nonrel eq 3}), even in the absence of EP violation, the real and imaginary parts of the mass do not couple to the Newtonian potential in the same way, which is a remarkable consequence of our formalism. Additionally, our derivation indicates limitations in the 
class of experiments where our treatment would be valid, and also suggests the path to describe other conceivable experiments.

Regarding the experimental perspectives of our model, we have presented detailed predictions appropriate for a COW-type experiment, along with definite signals to be searched for. At first sight it seems that the sensitivity of current interferometry experiments would not be sufficient to look for significant bounds for EP violations by unstable particles. However, there are some experiments, which may be feasible in the near future, having the sensitivity required to set nontrivial constraints on our parameter $\xi$. We also want to emphasize that our treatment of the COW experiment suggests an alternative resolution to the issue of the disappearance of the interference pattern as the tilt angle increases.

As a final thought we remark that the empirical tests suggested by this paper lie at the heart of the interplay of quantum mechanics and general relativity. The fact is that unstable particles can only be described as superposition of states, something intimately tied with the quantum aspects of nature, while the EP is one of the pillars of our understanding of gravitation. Moreover, we should point out that, at present, there is no compelling experimental evidence that can be taken as guidance of how gravity is supposed to interact with a system in quantum superposition. Thus, we believe that the experiments we propose represent interesting candidates from which we can obtain empirical clues as to how these two aspects of nature interact with each other.

\subsection*{Acknowledgments}
The authors wish to thank Roberto Colella and Sam Werner for many helpful discussions. This work was supported in part by the CONACYT grants 101712 and 103486, UNAM-PAPIIT grant IN107412, the Department of Energy grant DE-FG02-91ER40661, and by the Indiana University Center for Spacetime Symmetries.

\appendix

\section{Propagation of wavepackets in a COW experiment}\label{appendix COW}

\begin{figure}
\centering
\includegraphics[width=0.9\textwidth]{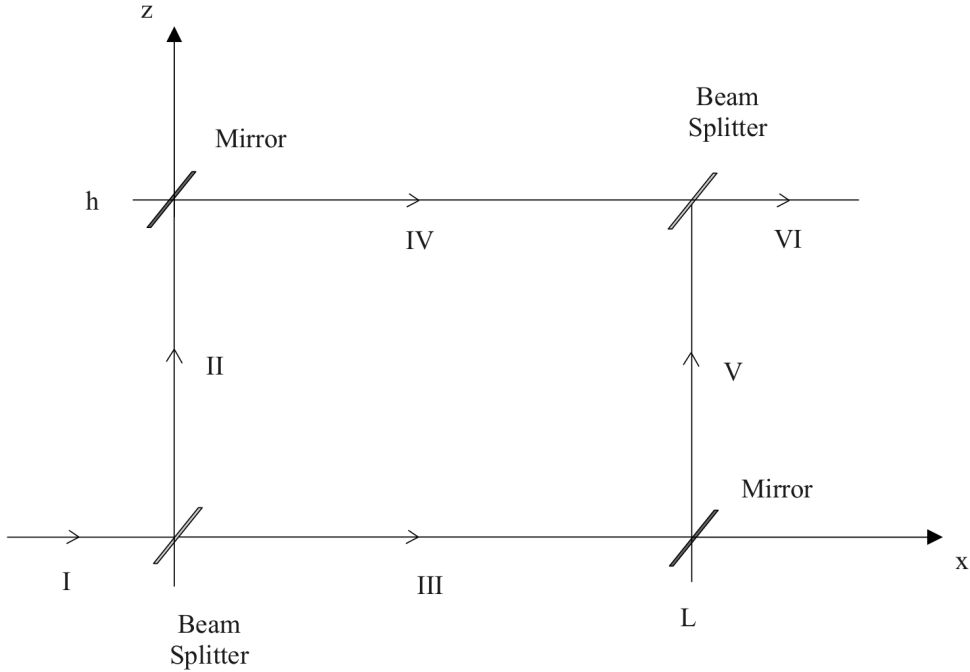}
\caption{\label{COWfigure}The COW setting under consideration where the coordinate system and the labels for the different segments are shown.}
\end{figure}

In this Appendix we study the propagation of a wavepacket in a COW configuration according to Eq.~(\ref{nonrel eq 4}). We use a reference frame adapted to the experimental setup where $x$ and $z$ are the horizontal and vertical coordinates, respectively (see Fig.~\ref{COWfigure}). As the Newtonian potential is constant in the horizontal segments of the interferometer, the wave-functions in these segments can be written as superpositions of solutions of the form
\begin{equation}\label{ansatz}
\phi(t,\vec{x})=e^{i\tilde{p}x-iEt},
\end{equation}
where $E$ is a real constant. Note, however, that as discussed in the text, the above expression cannot be considered as a Fourier transform on both space and time coordinates, something that can be seen explicitly in the fact that $\tilde{p}$ is complex; it is just an ansatz for the solution of Eq.~(\ref{nonrel eq 4}). (Bear in mind that we are not deriving the nonrelativistic equation but searching for its solutions.) By inserting our ansatz (\ref{ansatz}) into Eq.~(\ref{nonrel eq 4}) we obtain
\begin{eqnarray}
\tilde{p}^2&=&\frac{2m}{(1+U)^2}\left[\left(E-m-mU-m\frac{U^2}{2}-\Gamma U\xi^I\right)+i\Gamma\left(U(1+\xi^R)+\frac{1}{2}+\frac{U^2}{2}\right)\right],\label{p complex}
\end{eqnarray}
where we have written $\xi=\xi^R+i\xi^I$, with $\xi^R,\xi^I\in \mathbb{R}$. The solution of Eq.~(\ref{nonrel eq 4}) in the vertical segments, $ \tau(z)$, can be explicitly obtained; however, as it appears in both arms of the interferometer, it is not relevant for this calculation. 

Our strategy is to first find how a plane-wave basis propagates through the COW interferometer, which is later used to form Gaussian wavepackets. In order to do so, we study the effect of each beam splitter and mirror by solving a flux conservation equation. We start with the first beam splitter which is located at the origin of the reference frame. If we denote by $I$ the initial beam, $II$ the vertical segment at the first beam splitter and $III$ the beam that passes the splitter in the original direction (see Fig.~\ref{COWfigure}), then the associated wavefunctions are
\begin{eqnarray}
\label{phiI}\phi_I(x,t)&=&B_I e^{i\tilde{p}_lx-iEt},\\
\label{phiII}\phi_{II}(x,t)&=& C \tau(z)e^{-iEt},\\
\label{phiIII}\phi_{III}(x,t)&=&B_{III}e^{i\tilde{p}_lx-iEt},
\end{eqnarray}
where $\tilde{p}_l$ corresponds to $\tilde{p}$ as defined in expression (\ref{p complex}) for the Newtonian potential in the lower horizontal segments, and $B_{I}$, $B_{III}$ and $C$ are normalization constants. At the beam splitter the flux conservation equation implies
\begin{equation}\label{flux conserv}
\left| \vec{j}_I \right|=\left| \vec{j}_{II}\right|+\left|\vec{j}_{III} \right|,
\end{equation}
where $ \vec{j}_I =\Im(\phi_I^* \nabla\phi_I)/m$ and there are corresponding definitions for the other wavefunctions. (Here and subsequently, $\Re$ and $\Im$ respectively denote the real and imaginary parts of the argument.) Observe that the wavefunction is discontinuous at the beam splitter; this is due to the approximation that beams run over a line and because we are neglecting the splitters width. If we assume that the splitters divide the beams into two equal beams we get $\vec{j}_{III} = \vec{j}_{I}/2$. Normalizing $\phi_I(x,t)$ so that at the beam splitter $|\vec{j}_{I}|=1$, we obtain
\begin{eqnarray}\label{flux conserv3}
\frac{1}{2}=\frac{|C|^2}{m}\Im(\tau^*(0)\tau'(0))=\frac{|C|^2}{m}\Im(\tau'(0)),
\end{eqnarray}
where we use Eq.~(\ref{phiII}). In addition, we consider that the phase in the two resulting beams is equal to the phase in the incident beam, since the splitters add the same phase to the two beams. Therefore, if we suppose that $B_I=|B_I|$, we have $B_{III}=B_{I}/\sqrt{2}$ and $C=|C|$. 

We first compute the evolution through the upper segment. We label $IV$ the upper horizontal segment of the arrangement (as shown in Fig.~\ref{COWfigure}) and we assume
\begin{equation}\label{psi IV}
\phi_{IV}=B_{IV} e^{i\tilde{p}_ux-iEt},
\end{equation}
where $\tilde{p}_u$ means $\tilde{p}$ at the upper segment. In this case the flux conservation equation is $\left| \vec{j}_{II} \right|=\left| \vec{j}_{IV}\right|$, which implies
\begin{equation} \label{BIV}
C^2 \Im (\tau^*(h) \tau'(h))=|B_{IV}|^2\Re (\tilde{p}_u),
\end{equation}
where we use the fact that this mirror is located at the point $x=0$, $z=h$. Assuming that the phases of the incoming and outgoing wavefunctions coincide, which is justified since the two mirrors produce the same phase shift, we find
\begin{equation}
\phi_{IV}=\sqrt{\frac{m\Im (\tau^*(h) \tau'(h))}{2\Im(\tau'(0))\Re (\tilde{p}_u)}} \text{phase}(\tau(h)) e^{i\tilde{p}_u x-iEt},
\end{equation}
where we utilize the relations (\ref{flux conserv3}) and (\ref{BIV}).

We now turn to the lower path. Labeling $V$ the vertical segment at $x=L$, we can write
\begin{equation}\label{psi V}
\phi_{V}=D \tau(z) e^{-iEt},
\end{equation}
where $D$ is a normalization constant. In this case the flux conservation equation at the lower mirror, located at $x=L$ and $z=0$, can be written as 
\begin{equation}
|D|=|B_{III}| e^{-\Im (\tilde{p}_l)L} \sqrt{\frac{\Re (\tilde{p}_l)}{\Im (\tau'(0))}}.
\end{equation}
Moreover, by comparing the phase of $\phi_{III}$ and $\phi_V$ at this point we can check that the phase of $D$ is $e^{i\Re(\tilde{p}_l) L}$. Therefore
\begin{equation}\label{psi V 2}
\phi_{V}= e^{i \tilde{p}_l L-iEt} \sqrt{\frac{m}{2\Im (\tau'(0))}} \tau(z),
\end{equation}
where we use $B_{III}=\sqrt{m/2\Re( \tilde{p}_l) }$.

At the upper beam splitter, located at $x=L$ and $z=h$, we have to combine the wavefunctions $\phi_{IV}$ and $\phi_V$ into a new horizontal wavefunction $\Phi_E$. However, we must first consider the effect of the beam splitter over $\phi_V$ which changes the flux direction into the $x$ axis. Denoting $\phi_{VI}=B_{VI}e^{i\tilde{p}_ux-iEt}$ as the horizontal wavefunction produced from $\phi_V$ by the rotation, the condition $|\vec{j}_V|=|\vec{j}_{VI}|$ implies
\begin{equation}
|B_{VI}|=e^{\Im(\tilde{p}_u-\tilde{p}_l)L} \sqrt{\frac{m}{2\Re(\tilde{p}_u) }}\sqrt{\frac{\Im(\tau^*(h) \tau'(h))}{\Im (\tau'(0))}}.
\end{equation}
The fact that we neglect the phase shift due to the splitter allows us to set 
\begin{equation}
B_{VI}=|B_{VI}|e^{i\Re(\tilde{p}_l-\tilde{p}_u) L}\text{phase}(\tau(h)),
\end{equation}
and thus
\begin{eqnarray}
\phi_{VI}&=& \sqrt{\frac{m}{2\Re(\tilde{p}_u) }}\sqrt{\frac{\Im(\tau^*(h) \tau'(h))}{\Im (\tau'(0))}}e^{i(\tilde{p}_l-\tilde{p}_u) L}\text{phase}(\tau(h))e^{i\tilde{p}_ux-iEt}.
\end{eqnarray}
With this result we compute
\begin{equation}
\Phi_E=\phi_{IV}+\phi_{VI}=\Lambda e^{-iEt}e^{i\tilde{p}_u x}\left( 1+e^{-i \widetilde{\Delta p} L}\right).
\end{equation}
where we define
\begin{equation}
\Lambda=\sqrt{\frac{m\Im (\tau^*(h) \tau'(h))}{2\Im(\tau'(0))\Re (\tilde{p}_u)}} \text{phase}(\tau(h)),
\end{equation}
and $\widetilde{\Delta p} = \tilde{p}_u-\tilde{p}_l$.

We want to find how a wavepacket with energy centered at $E_0$ propagates through this arrangement. This wavepacket can be described by the wavefunction
\begin{equation}
\Phi=\int\frac{F(E)}{\Lambda} \Phi_E dE,
\end{equation}
where $F(E)/\Lambda$ is a function peaked at $E_0$ (the factor $1/\Lambda$ is conventional). If the detector is placed at $x=L$, namely, next to the upper beam splitter, and assuming that it captures particles for a long time (which is approximated as infinite), the nonrelativistic probability for the detector at $x=L$ to capture a particle is
\begin{eqnarray}
P(L)&=&\int_{-\infty}^\infty|\Phi(L,t)|^2dt\nonumber\\
&=& 2\pi \int |F(E)|^2 e^{-2\Im(\tilde{p}_u) L}\left[1+e^{2\Im(\widetilde{\Delta p}) L} +2 \cos(\Re(\widetilde{\Delta p}) L)e^{\Im(\widetilde{\Delta p}) L}\right]dE.\label{P}\end{eqnarray}
As in Eq.~(\ref{IntegratedProb}), we use the fact that the integral over $t$ gives a Dirac delta function to integrate $E'$. Note that the interference involves only components with the same energy, as it is expected. We write the Gaussian wavepacket centered at $E_0$, and having width $\Delta E=1/\Delta t$, as
\begin{equation}
F(E)=\left(\sqrt{\frac{2}{\pi}}\frac{\Delta t}{N_{\Delta}}\right)^{1/2} e^{-\frac{1}{4}\Delta t^2
(E-E_0)^2},\label{gaus}
\end{equation}
where $N_\Delta$ is such that 
\begin{equation}
\int|F(E)|^2 dE=1.\label{fnor} 
\end{equation}
Defining $\epsilon=\Delta t(E-E_0)$, Eq.~(\ref{P}) reads, up to a normalization constant, as
\begin{equation}
P(L)=\int e^{-\frac{\epsilon^2}{2}}\left(e^{-\beta_u}+e^{-\beta_l}+2 e^{-\chi}\cos{\varphi}\right)d\epsilon,\label{Pfull}
\end{equation}
where we define
\begin{eqnarray}
\beta_{l,u}&=&2\Im(\tilde{p}_{l,u})L,\\
\chi&=&\Im\left(\tilde{p}_{l}+\tilde{p}_{u}\right)L,\\
\varphi&=&\Re{(\widetilde{\Delta p})}L.\label{def varphi}
\end{eqnarray}
In the approximation where the wavepackets are narrow, we can work to first order in $(E-E_0)/E_0=\epsilon/(E_0\Delta t)$. If we also neglect terms quadratic in $\Gamma/m$ and $U$, but retain those terms that go like $\Gamma U/m$, we obtain
\begin{eqnarray}
\beta_{l,u}&=&\frac{L \Gamma}{v_0^3}\left(U_{l,u}+v_0^2[1+U_{l,u}(1+2\xi^R)]\right) -\frac{ \epsilon L \Gamma}{2 v_0^3 E_0\Delta t}\left(3 U_{l,u}+v_0^2[1+U_{l,u}(1+2\xi^R)]\right), \label{beta approx}\\
\chi&=&\frac{L\Gamma }{2 v_0^3}\left(\langle U\rangle+v_0^2[1+\langle U\rangle(1+2\xi^R)]\right)-\frac{\epsilon L\Gamma }{m v_0^5E_0\Delta t}\left(3\langle U\rangle+v_0^2[1+\langle U\rangle(1+2\xi^R)]\right),\label{chi approx}\\
\varphi&=&\frac{mL \Delta U}{v_0}\left(1+\frac{\Gamma\xi^I}{m}+v_0^2\right)-\frac{\epsilon mL \Delta U}{2v_0^2E_0\Delta t}\left(1+\frac{\Gamma}{m} \xi^I- v_0^2\right).\label{varphi approx}
\end{eqnarray}
In Eqs.~(\ref{beta approx})--(\ref{varphi approx}) we have introduced the initial speed $v_0=\sqrt{2E_0/m}$, average potential $\langle U\rangle=(U_u+U_l)/2$ and potential difference $\Delta U=U_u-U_l$. At this point it is possible to observe that in the limit where $\Gamma=0$, $\Delta E = 1/\Delta t =0$ and $v_0\ll 1$, we recover the usual expressions for a COW experiment. Using these limits along with $U = gz$ and $h = h_{0}\cos\theta$, we find
\begin{eqnarray}
\beta_{l,u}&=& 0, \\
\chi&=& 0, \\
\varphi & = & \frac{mgh_{0}L\cos\theta}{v_{0}} \equiv \varphi_{\rm{COW}}.
\end{eqnarray}
By inserting Eqs.~(\ref{beta approx})--(\ref{varphi approx}) into Eq.~(\ref{Pfull}), the explicit integral can be computed. Its full expression is too long to be written explicitly, but it has been used to make the plots shown in Figs.~\ref{fig:fPi} of $P(L)$ as a function of the angle $\theta$ between the COW apparatus and the gravitational potential gradient, and for different values of $\xi$. 

\section{Limiting cases}\label{limitcases}
 
In this Appendix we calculate the probability of detecting a particle in the COW configuration of Appendix~\ref{appendix COW} where the initial energy wavepackets are extremely narrow or wide. This is done to fill the gaps left in our previous calculation and to test if the results match our intuition. We expect that, when the state has a well defined energy (and it is widely distributed in space), the probability (\ref{Pfull}) reduces to the typical interference pattern between two plane waves times a suitable decaying factor. On the other hand, when the initial state is highly localized (in space), the interference pattern should disappear as the wavepacket components traveling in the two interferometer segments do not arrive at the detector simultaneously. In the rest of the Appendix we focus in the interference term contribution to the probability, which is given, up to a normalization constant, by the third term in the full detection probability (\ref{Pfull}), namely,
\begin{equation}
P_{\rm{int}}(L)=\int e^{-\frac{\epsilon^2}{2}}e^{-\chi}\cos{\varphi}d\epsilon.\label{Pint}
\end{equation}

\subsection*{Small energy uncertainty}

This limit can be easily obtained from the last part of Appendix~\ref{appendix COW} by taking $\Delta E = 1/\Delta t \rightarrow 0$ in Eqs,~(\ref{varphi approx})--(\ref{chi approx}). We find
\begin{eqnarray}
\chi&=&\frac{L\Gamma }{2 v_0^3}\left(\langle U\rangle+v_0^2[1+\langle U\rangle(1+2\xi^R)]\right),\\
\varphi&=&\frac{mL \Delta U}{v_0}\left(1+\frac{\Gamma\xi^I}{m}+v_0^2\right)= \varphi_{\rm{COW}}\left(1+\frac{\Gamma\xi^I}{m}+v_0^2\right),
\end{eqnarray}
which are independent of $\epsilon$. Thus, it can be easily checked that the interference term contribution to the probability corresponds to the typical interference term between two plane waves times a $\Gamma$-dependent decaying exponential, as expected.

\subsection*{Large energy uncertainty}

We define $\delta t$ as the difference between the time at which the center of the wavepackets traveling along the upper and lower arms of the interferometer arrive at the detector. In order to determine $\delta t$ we use the stationary phase approximation, that is, given the wavepacket
\begin{equation}
\Phi_i(x,t)= \int F(E)
e^{i\vartheta_i(x,t;\:E)}dE,\label{wpack}
\end{equation}
we define its center as the spatial point $x_{m}(t)$ where the amplitude of the above expression reaches its maximum. This occurs when $d\vartheta_i/dE$ around $E_0$ is small, since under such conditions there is constructive interference. If we define the phases of the upper and lower wavepacket as
\begin{eqnarray}
\vartheta_l&=&\Re{(\tilde{p}_u)} x-\Re{(\widetilde{\Delta p})L}-Et,\label{phl}\\
\vartheta_u&=&\Re{(\tilde{p}_u)} x-Et,\label{phu}
\end{eqnarray}
it is not difficult to see that, at $x=L$,
\begin{equation}
\delta t=\frac{\partial}{\partial E}\left.\left(L\Re{(\widetilde{\Delta p})}\right)\right|_{E_0}
=\left.\frac{\partial \varphi}{\partial E}\right|_{E_0},\label{dt}
\end{equation}
where in the last step we use the definition of $\varphi$ given in Eq.~(\ref{def varphi}).

\begin{figure}
\centering
\includegraphics[width=0.5\textwidth]{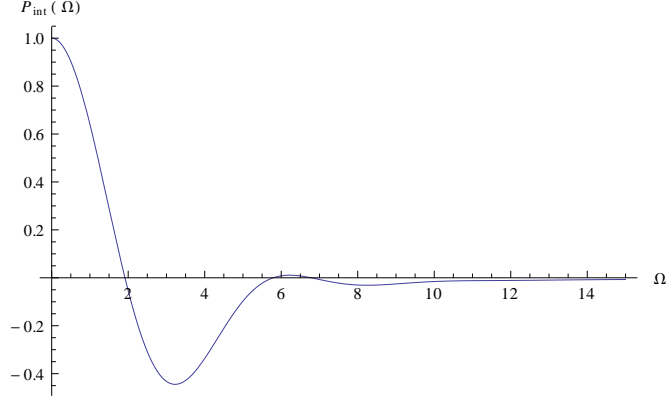}
\caption{\label{fig:Pint 2}The interference term contribution to the probability $P_{\rm{int}}$ as a function of $\Omega$ with $\Gamma=0$. It can be observed that for $\Omega \geq 10$ the probability of interference tends to zero.}
\end{figure}

It can be explicitly checked that in the limit $\Delta E = 1/\Delta t \rightarrow \infty$ the expressions for $\varphi$ and $\chi$ take the form
\begin{eqnarray}
\chi&=&v_0 \Gamma L\left(1+\langle U\rangle(1+2\xi^R)\right)\sqrt{\frac{\epsilon}{E_0\Delta t}},\label{chi0}\\
\varphi&=& \Omega \sqrt{\epsilon},\label{vphi0}
\end{eqnarray}
where we define
\begin{equation}
\Omega = \frac{2v_0^2}{v_0^2-1-\xi^{I}\Gamma/m}\sqrt{\frac{E_0}{\Delta t}}. 
\end{equation}
Thus, the interference term contribution to the probability becomes
\begin{equation}
P_{\rm{int}}= \int
e^{-\frac{1}{2}\epsilon^2} e^{-\chi}
\cos{\left(\Omega\sqrt{\epsilon}\right)} d\epsilon.\label{Pint0}
\end{equation}
Observe that if $\Delta t\rightarrow 0$ then $\Omega\rightarrow \infty$, and therefore Eq.~(\ref{Pint0}) is the integral of a rapidly oscillating cosine with a decreasing exponential envelope which, as it is well known, vanishes. Note also that $\chi$ as given in Eq.~(\ref{chi0}) is a positive definite function and thus, the previous argument applies for any value of $\Gamma$. In order to show that the interference pattern is erased when $\delta t\gg\Delta t$, we fix $\Gamma=0$ in Fig.~\ref{fig:Pint 2} where the dependence of $P_{\rm{int}}$ on $\Omega$ is shown. This plot shows that the interference pattern vanishes for large values of $\Omega$ which, in turn, occurs when $\Delta t\rightarrow 0$. A similar effect is found in Ref.~\cite{Zych} where the loss of interference is interpreted as a measure of both relativistic time-dilation and the complementary principle.

\end{document}